# Bonded Thin Film Lithium Niobate Modulator on a Silicon Photonics Platform Exceeding 100 GHz Bandwidth

**Authors:** Peter O. Weigel[1,*], Jie Zhao[1], Kelvin Fang[1], Hasan Al-Rubaye[1], Douglas Trotter[2], Dana Hood[2], John Mudrick[2], Christina Dallo[2], Andrew T. Pomerene[2], Andrew L. Starbuck[2], Christopher T. DeRose[2], Anthony L. Lentine[2], Gabriel Rebeiz[1] and Shayan Mookherjea[1,*]

Affiliations:

[1] University of California, San Diego, Electrical & Computer Engineering, La Jolla, California 92093-0407, USA

[2] Sandia National Laboratories, Applied Microphotonic Systems, Albuquerque, New Mexico, 87185, USA

* Correspondence to: pweigel@eng.ucsd.edu, smookherjea@ucsd.edu

**Electro-optic modulation, the imprinting of a radio-frequency (RF) waveform on an optical carrier, is one of the most important photonics functions, being crucial for high-bandwidth signal generation, optical switching, waveform shaping, data communications, ultrafast measurements, sampling, timing and ranging, and RF photonics. Although silicon (Si) photonic electro-optic modulators (EOMs) can be fabricated using wafer-scale technology compatible with the semiconductor industry, such devices do not exceed an electrical 3-dB bandwidth of about 50 GHz** [1]**, whereas many applications require higher RF frequencies. Bulk Lithium Niobate (LN)** [2] **and etched LN modulators** [3] **can scale to higher bandwidths, but are not integrated with the Si photonics fabrication process adopted widely over the last decade. As an**

**alternative, an ultra-high-bandwidth Mach-Zehnder EOM based on Si photonics is shown, made using conventional lithography and wafer-scale fabrication, bonded to an unpatterned LN thin film. This hybrid LN-Si MZM achieves beyond 100 GHz 3-dB electrical bandwidth. Our design integrates silicon photonics light input/output and optical components, including directional couplers, low-radius bends, and path-length difference segments, realized in a foundry Si photonics process** [4]. **The use of a simple low-temperature (200°C) back-end integration process to bond a postage-stamp-sized piece of LN where desired, and achieving light routing into and out of LN to harness its electro-optic property without any etching or patterning of the LN film, may be broadly-useful strategies for advanced integrated opto-electronic microchips.**

Mach-Zehnder modulators (MZM) consist of light being split equally into two pathways, each experiencing a voltage-driven optical phase shift, typically between 0 and $\pi/2$ radians in each arm with opposing signs, and recombining to result in controllable optical transmission. The MZM structure is one of the most widely-used EOM's in practical applications, and LN is a suitable and tested material for high-speed and high-bandwidth EOM, with stand-alone MZM devices having reached 110 GHz optical (70 GHz electrical) 3-dB bandwidth about two decades ago in an unpackaged device [2]. However, the traditional design and fabrication approach of LN MZM's, based on ion exchange or implantation into bulk LN, is a relatively slow, expensive and labor-intensive process, which is not compatible with the complex, multi-functional integrated optics microchips being currently developed and deployed. As a step towards integrated modulators, a variety of approaches using thin-film LN [5, 6] have been reported [7, 8, 9, 10, 11]. The ability to match the optical and microwave indices by varying the dimensions of the LN layer and the rib-loading Si waveguide offer new design opportunities to achieve true optical-RF phase matching to very high frequencies without artificial velocity matching structures. Indeed, electro-optic modulation sidebands have been measured to several hundred gigahertz (though not the 3-dB point) for ridge waveguide LN EOMs [12]. The combination of beyond-100 GHz electrical (3-dB) bandwidth and a design / fabrication approach and compatible materials (e.g., oxide, rather than polymeric, bonding layers, and aluminum, rather than gold, electrodes) that is compatible with, and leverages, wafer-scale and foundry-based approaches popularized by Si photonics would be a significant advancement from both a performance and a practical (e.g., fabrication, cost, scalability) perspective.

In our fabrication approach, depicted in Fig. 1, MZM's were built on a silicon photonics platform, using photolithography on silicon-on-insulator wafers (220 nm Si thickness, 3 μm oxide thickness) and did not require sub-resolution features unlike most plasmonic or polymeric slot modulators [13, 14]. Silicon thinning (down to 150 nm) and feature patterning were followed by oxide deposition and subsequent chemical mechanical polishing and oxide thinning by a timed wet etch (diluted hydrofluoric acid) process. After die segmentation, commercially-procured x-cut thin-film LN on insulator (NanoLN, Jinan Jingzheng Electronics Co. Ltd.) was bonded over a large area (~1 cm$^2$) with a pressure of 45 kPa, but not processed further (e.g., no etching [9, 15] or sawing [16] of LN was performed). Oxide bonding was done at room temperature after surface cleaning and surface plasma activation steps. The bonded sample was thermally annealed at 200°C for one hour under pressure. The bonded stack has been shown to withstand repeated temperature-cycling to at least 300°C [17], sufficient for the post-processing required here. In fact, several fabricated chips were repeatedly processed, after bonding, through multiple cycles of electrode formation, removal and re-formation, in search of the optimal dimensions. No debonding or noticeable degradation to the stability or quality of the samples occurred during these additional process steps. The LN die handle was removed, followed by coplanar waveguide electrode formation using aluminum deposition, with total electrode thickness of 1.6 μm. A fully fabricated chip is shown in Fig. 1c; a microscope image of the EOM is provided in Fig. 1d. The electrodes used here are more than 15 times thinner than those used in Ref. [2], thus improving fabrication practicality. The silicon photonic features were made on a high-resistivity Si handle wafer (measured resistivity of the handle after HF etch to remove native oxide is around $6 \times 10^3$ Ω.cm), potentially mitigating piezoelectric resonances from traditional LN substrates [18]. As described below, the optical input and output from the MZM section were through (crystalline) silicon photonic waveguiding structures.

For the MZM reported here, the silicon photonics region outside the bonded LN area included four types of optical waveguide structures: fully-etched tapers for light input and output from the chip (edge couplers), single-mode broadband directional couplers (>15 dB extinction ratio, ER, throughout 1525 nm -1575 nm, maximum ER of approximately 30 dB), path-length difference (PLD) segment (including spline curve bends), and adiabatic waveguide tapers for inter-layer transitions (Si-to-LN and vice-versa). Precise foundry processing of the Si photonic features results in accurate and repeatable formation of the directional coupler splitting ratio. Since the LN layer is neither patterned nor etched in our design, there was no alignment issue at the bonding step; the Si features alone determine the optical propagation path.

Adiabatic waveguide tapers were designed to achieve a vertical inter-layer transition (from Si to LN, and the reverse). As shown in Fig. 2, the design uses the TE-polarized fundamental guided mode, which is also used in conventional silicon photonics at 1.5 μm wavelengths [19]. Since the refractive index of Si at these wavelengths (approximately 3.5) is significantly higher than that of LN (approximately 2.2), the large index difference enables control of the mode size and location (i.e., mainly in the Si rib or the LN slab) through lithography of the Si layer alone. Thus, only the width of the Si waveguide (w) was varied in our design; when w > 600 nm, light at 1.55 μm is mostly confined within the Si rib with confinement factor $\Gamma_{Si}$ = 64% (Mode A) and $\Gamma_{Si}$ = 58% (Mode B). For w = 320 nm, light is guided in Mode C and "sees" the LN slab layer, with confinement fraction in the LN layer calculated as $\Gamma_{LN}$ = 81% and $\Gamma_{Si}$ = 5%. Longitudinal Poynting vector simulations of these modes are shown in Fig. 2c. We do not reduce the width of the Si rib (w) in Mode C further, in order to have a laterally confined optical mode despite the high value of $\Gamma_{LN}$, which does not experience high optical loss from the metal electrodes. Also, by not letting the mode expand further in width, the Mode B – Mode C transition loss is kept low, and is estimated as 0.1 dB from simulations, and is described in more detail in previous work [20]. A benefit of these high-bandwidth modulators is that $\Gamma_{LN}$ has less variation with small errors in fabricated waveguide dimensions than plasmonic or polymeric slot waveguide MZM, and the minimum feature sizes are easily achieved by today's silicon photonics processing technology.

Vertical, inter-layer transitions to and from the hybrid LN-Si region occur only where needed, inside the perimeter of the bonded region. Optical losses between Modes A and B are minimized by keeping the Si waveguide wide (w = 650 nm) when crossing into the hybrid region. Thus, the edges of the bonded thin film, even if rough on the scale of the optical wavelength, do not significantly affect optical propagation. This makes the back-end integration of thin-film LN simple and feasible, without requiring precision etching or patterning of either LN or silicon after bonding. Complex waveguiding circuits can be built up with a single bonded layer and multiple vertical transitions, as shown elsewhere [20], but were not required here.

A design library of hybrid LN-Si components was created to aid in simulations and design within the Lumerical Interconnect simulation environment [21]. Light input and output was achieved using tapered single-mode, polarization-maintaining fibers. From test structures, an optical propagation loss of -0.6 dB/cm in the hybrid LN-Si region was measured. The propagation losses in the Si-only regions were about -1.3 dB/cm and are kept short in this design. The edges of the silicon photonic chip were lightly polished, but not fully prepared or packaged; hence, the edge coupling loss was about -3 dB per edge and the total fiber-to-fiber insertion loss was -13.6 dB. The calculated intrinsic loss of the full MZM (not including edge couplers), based on the measured propagation loss (-0.6 dB/cm in the EOM, -1.3 dB/cm outside the EOM) and the device length (0.5 cm for the EOM, 1.67 cm outside the EOM), simulated inter-layer transition loss estimates (-0.1 dB each), should be about -2.9 dB. However, the actual insertion loss was around -7.6 dB. The additional loss (4.7 dB) is likely due to higher-than-expected attenuation in the broadband directional couplers and tapers, as well as an estimated (from simulations) additional -0.4 dB loss due to the electrical lines, which pass directly over the optical mode (see Fig. 1d). It is also possible that the propagation losses in this particular device were higher than expected due to the multiple re-fabrication performed on this chip in search of the optimum electrode structure. At low speeds, the MZM demonstrated a high extinction ratio (> 20 dB) as shown in Fig. 3a, with $V_\pi L$ = 6.7 V.cm at DC for an L = 0.5 cm device.

RF measurements were performed on a bare-die chip using 50-$\Omega$ probes rated to 110 GHz and using laboratory equipment and RF waveguide components also rated and calibrated to about 110 GHz. The RF driving waveform was either from an RF oscillator (up to 67 GHz) or three different frequency multipliers (sequentially covering the range of frequencies up to 106 GHz). GSG probes were used for both launch and termination. Calibration of the signal pathway was performed using a high-frequency RF power sensor. To inform a computational model of the expected behavior, electrical S-parameters were measured using a Vector Network Analyzer up to 110 GHz, as shown in Fig. 3b, and analyzed using standard algebraic transformations and lossy transmission line circuit analysis [22]. The artifact near 90 GHz is attributed to repeated testing by contact probes, and does not disrupt the smoothly-varying general trend seen in the data. The microwave refractive index, $n_m$, was fitted to a power-law equation and the characteristic impedance, $Z_c$, was fitted to a first-order polynomial of the RF frequency. As shown in Fig. 3c, $n_m$ = 2.25 and $Z_C$ varied between 53.4 and 55.1 $\Omega$ from dc to 110 GHz. The microwave loss, $\alpha_m$, when fitted to a power-law equation, followed approximately an $f^{1/4}$ dependence, in contrast with the typical $f^{1/2}$ dependence in traditional EOMs [2]. Because this device has relatively thin electrodes and a silicon substrate, $\alpha_m$ is not due solely to conductor loss, and instead, includes a combination of conductor, substrate, and radiation losses.

The small-signal optical modulation response of an electro-optic Mach-Zehnder modulator based on phase modulation (a good approximation of the Pockels effect in LN) is [23]:

$$m(\omega) = \frac{R_L+R_G}{R_L} \left|\frac{Z_{in}}{Z_{in}+Z_G}\right| \left|\frac{(Z_L+Z_0)F_{u_+}+(Z_L-Z_0)F_{u_-}}{(Z_L+Z_0)\exp(\gamma_m L)+(Z_L-Z_0)\exp(-\gamma_m L)}\right| \quad (1)$$

where $n_o$ is the optical group index, $n_m$ and $\alpha_m$ are the microwave index and loss coefficient, L is the phase-shifter arm length, $\omega$ is the RF frequency, c is the speed of light, $Z_L$ ($R_L$) and $Z_G$ ($R_G$) are the load and generator impedances (and, respectively, resistances), $Z_0$ is the electrode transmission line characteristic impedance, $Z_{in}$ is the RF line input impedance, and using the definitions $F(u) = [1-\exp(u)]/u$; $u_\pm = \pm \alpha_m L + j(\omega/c)[\pm n_m - n_o]L$; and complex RF propagation constant $\gamma_m = \alpha_m + j \omega n_m/c$.

Equation (1) is a low-pass filter type response, whose 3-dB electrical roll-off frequency ($f_{3dB,el}$) is maximized by matching of the optical and RF indices, matching the load and generator impedances, and minimizing the RF loss, but is not affected by the optical propagation loss. Lower optical propagation loss ($\alpha_{opt}$) improves overall transmission, but for values of $\alpha_{opt} < 1$ dB/cm and device lengths < 1 cm, the actual measured losses are dominated by non-idealities, such as imperfect chip coupling, or non-unitary power splitting at directional couplers, and further optical loss reduction plays only a minor role.

The implications of Equation (1) can be understood by assessing its behavior with regard to each significant parameter in turn. Assuming velocity and impedance matching, the RF-loss limited bandwidth results in a 3-dB point of $\alpha_m(f_{3dB,el}).L = 6.4$ dB. Using L = 0.5 cm for the device under test, and requiring the 3-dB electrical frequency $f_{3dB,el} \geq 100$ GHz, we need $\alpha_m(100 \text{ GHz}) \leq 12.8$ dB.cm$^{-1}$. As shown in Fig. 3, measurements showed $\alpha_m(100 \text{ GHz}) = 7.7$ dB.cm$^{-1}$, well under the limit, and thus, RF losses are not a limitation. Assuming impedance matching and no RF loss, the bandwidth limitation $f_{3dB,el}.L = (0.13/\Delta n)$ GHz.m, where $\Delta n = n_m - n_o$. Thus, achieving $f_{3dB,el} \geq 100$ GHz for a 0.5 cm long device requires $\Delta n_{mo} = n_m - n_o \leq 0.26$. Calculations for our device indicate that $\Delta n_{mo} \approx 0.07$, which is well under the threshold for achieving 100 GHz electrical bandwidth. Moreover, the electrode length times bandwidth product [24] "LB" = $c/\Delta n$ = 428 GHz.cm (i.e., zero modulation response at f = 428 GHz for a 1 cm long device) whereas the fabricated device which achieves beyond 106 GHz modulation has L = 0.5 cm, a factor of two shorter. Thus, neither RF losses nor index matching are fundamental limitations, as they have been in the past.

The method of Ref. [25] was used to detect signals and modulation sidebands at an optical wavelength of 1560 nm. With the modulator biased at quadrature, the difference (log scale) between the optical intensity of the first sideband and carrier signal was used to extract the modulation index, and thus the frequency response, from 106 GHz down to 2 GHz (providing a safe margin for the 0.18 GHz resolution of the OSA). The peak-to-peak RF drive amplitude was about 1 Volt. Frequency multipliers were used for the frequency range above 67 GHz up to 106 GHz. The electro-optic response is shown in Fig. 4, and an (electrical) 3-dB bandwidth was observed to lie well beyond 106 GHz, the limit of our measurement capabilities. The measurement matches well with the calculated response (solid black line in Fig. 4), which was obtained by plugging the measured values of Fig. 3c into Eq. (1). Electro-optic measurements at frequencies beyond 67 GHz require RF multipliers and time-consuming calibrations using appropriate waveguides, cables, probes and detectors for each frequency band, as observed elsewhere [26]. Some of the scatter in the measurements at the highest frequencies arises from the calibration of the frequency extenders, which have nonlinear and discontinuous dispersion curves. The measured flat-spectrum modulation response is consistent with our simulation based on electrical S-parameter measurements, which predicts flat frequency response to even higher frequencies.

The product $V_\pi L$ = 6.7 V.cm (with a 0.5 cm length) in this device compares favorably with other high-bandwidth LN MZI's and commercial technology, but the present devices were not designed to minimize $V_\pi L$. Recently, a fully-etched thin-film LN-on-SiO$_2$ MZM was reported, which achieved low $V_\pi.L$ = 2.2 V.cm [3]. In contrast with our device, etched LN structures have been shown as stand-alone devices, not integrated with other integrated photonics components, and etching LN may lead to concerns such as heat and pyroelectric charge buildup, structural defect formation and Nb depletion [27, 28]. Lower $V_\pi L$, though desirable, is arguably not the primary figure-of-merit of LN modulators, since an order-of-magnitude lower $V_\pi.L$ product can be achieved using organic [29], plasmonic [30] or graphene [31] modulators, which, have not demonstrated such wide 3-dB electrical bandwidths, and are less widely adopted than LN modulators in practical usage.

Here, an un-patterned thin-film of LN was simply bonded at room temperature to the patterned and planarized Si waveguide circuits, with an anneal step at 200°C. In contrast, fabrication of bulk titanium-indiffused LN modulators and doped III-V or Si modulators require at least 600°C and typically 900-1000°C [32, 33], limiting them to either standalone or front-end-of-line device fabrication. Based on our approach, integration of electronic circuits alongside a very-high-bandwidth LN EOM may be envisioned. More complex integrated optics can be realized, which use a wider range of silicon photonic components such as filters, interferometers and detectors alongside one or several ultra-high-bandwidth modulators, formed using single-step bonding. Integration may also avoid the challenges of traditional packaging of LN EOM's: the 70-GHz unpackaged LN modulator of Ref. [2] was reported to achieve a 3-dB bandwidth, when packaged into a stand-alone module, of only about 30 GHz [34].

In summary, we report an electro-optic Mach Zehnder modulator (MZM) based on single-mode silicon (Si) photonic circuits bonded at low temperature to an unpatterned, un-etched thin-film of lithium niobate (LN), thus utilizing the well-known Pockels electro-optic effect in only the desired section of the light pathway. Both theory and measurements support the performance of this device as a greater-than-100-GHz electrical bandwidth EOM, realized using a new design and fabrication process that brings lithium niobate, the traditional electro-optic material-of-choice in the first few decades of integrated optics, into compatibility with silicon photonics, the more recent platform for more complex integrated optics. The input and output are in silicon photonics and, through the use of inter-layer vertical waveguide transitions, the device is not sensitive to the rough edges, if any, of the LN thin film. The fabrication process, which does not require etching or sawing of LN, is based on a standard silicon photonics foundry fabrication flow. Such a device can bring ultrawide electro-optic bandwidths to integrated silicon photonics, and benefit applications in analog and digital communications, millimeter-wave instrumentation, analog-to-digital conversion, sensing, antenna remoting and phased arrays.

**References:**

[1] J. Sun, M. Sakib, J. Driscoll, R. Kumar, H. Jayatilleka, Y. Chetrit and H. Rong, *A 128 Gb/s PAM4 Silicon Microring Modulator,* San Diego, CA, 2018.

[2] K. Noguchi, O. Mitomi and H. Miyazawa, "Millimeter-Wave Ti:LiNbO3 Optical Modulators," *Journal of Lightwave Technology,* vol. 16, no. 4, pp. 615-619, 1998.

[3] M. Zhang, C. Wang, X. Chen, M. Bertrand, A. Shams-Ansari, S. Chandrasekhar, P. J. Winzer and M. Loncar, "Ultra-High Bandwidth Integrated Lithium Niobate Modulators with Record-Low Vpi," in *OFC*, San Diego, 2018.

[4] C. T. DeRose, M. Gehl, C. Long, N. Boynton, N. Martinez, A. Pomerene, A. Starbuck, C. Dallo, D. Hood, D. C. Trotter, P. Davids and A. Lentine, "Radio frequency silicon photonics at Sandia National Laboratories," in *2016 IEEE Avionics and Vehicle Fiber-Optics and Photonics Conference (AVFOP)*, Long Beach, CA, 2016.

[5] M. Levy, R. M. Osgood Jr., R. Liu, L. E. Cross, G. S. Cargill III, A. Kumar and H. Bakhru, "Fabrication of single-crystal lithium niobate by crystal ion slicing," *Applied Physics LEtters,* vol. 73, pp. 2293-2295, 1998.

[6] G. Poberaj, H. Hu, W. Sohler and P. Gunter, "Lithium niobate on insulator (LNOI) for micro-photonic devices," *Laser Photonics Review,* vol. 6, pp. 488-503, 2012.

[7] L. Chen, J. Chen, J. Nagy and R. M. Reano, "Highly linear ring modulator from hybrid silicon and lithium niobate," *Optics Express,* vol. 23, pp. 13255-13264, 2015.

[8] A. Rao, A. Patil, P. Rabiei, A. Honardoost, R. DeSalvo, A. Paolella and S. Fathpour, "High-performance and linear thin-film lithium niobate Mach-Zehnder modulators on silicon up to 50 GHz," *Optics Letters,* vol. 41, no. 24, pp. 5700-5703, 2016.

[9] A. J. Mercante, P. Yao, S. Shi, G. Schneider, J. Murakowski and D. W. Prather, "110 GHz CMOS Compoatible thin film LiNbO3 modulator on silicon," *Optics Express,* vol. 24, no. 14, pp. 15590-15595, 2016.

[10] C. Wang, M. Zhang, B. Stern, M. Lipson and M. Loncar, "Nanophotonic lithium niobate electro-optic modulators," *Optics Express,* vol. 26, no. 2, pp. 1547-1555, 2018.

[11] L. Chen, Q. Xu, M. G. Wood and R. M. Reano, "Hybrid silicon and lithium niobate electro-optical ring modulator," *Optica,* vol. 1, no. 2, pp. 112-118, 2014.

[12] J. Macario, P. Yao, S. Shi, A. Zablocki, C. Harrity, R. D. Martin, C. A. Schuetz and D. W. Prather, "Full spectrum millimeter-wave modulation," *Optics Express,* vol. 20, pp. 23623-23629, 2012.

[13] S. Zhu, G. Q. Lo and D. L. Kwong, "Phase modulation in horizontal metal-insulator-silicon-insulator-metal plasmonic waveguides," *Optics Express,* vol. 21, pp. 8320-8330, 2013.

[14] M. Ayata, Y. Fedoryshyn, W. Heni, B. Baeuerle, A. Josten, M. Zahner, U. Koch, Y. Salamin, C. Hoessbacher, C. Haffner, D. L. Elder, L. R. Dalton and J. Leuthod, "High-speed plasmonic modulator in a single metal layer," *Science,* vol. 358, no. 6363, pp. 630-632, 2017.

[15] G. Ulliac, V. Calero, A. Ndao, F. Baida and M. P. Bernal, "Argon plasma inductively coupled plasma reactive ion etching study for smooth sidewall thin film lithium niobate waveguide application," *Optical Materials,* vol. 53, pp. 1-5, 2016.

[16] N. Courjal, F. Devaux, A. Gerthoffer, C. Guyot, F. Henrot, A. Ndao and M. P. Bernal, "Low-loss LiNbO3 tapered-ridge waveguides made by optical-grade dicing," *Optics Express,* vol. 23, no. 11, pp. 13983-13990, 2015.

[17] P. O. Weigel and S. Mookherjea, "Reducing the thermal stress in a heterogeneous material stack for large-area hybrid optical silicon-lithium niobate waveguide micro-chips," *Optical Materials,* vol. 66, pp. 605-610, 2017.

[18] J. L. Nightingale, R. A. Becker, P. C. Willis and J. S. Vrhel, "Characterization of frequency dispersion in Ti-indiffused lithium niobate optical devices," *Applied Physics Letters,* vol. 51, pp. 716-718, 1987.

[19] Y. A. Vlasov and S. J. McNab, "Losses in single-mode silicon-on-insulator strip waveguides and bends," *Optics Express,* vol. 12, pp. 1622-1631, 2004.

[20] P. O. Weigel, M. Savanier, C. T. DeRose, A. T. Pomerene, L. A. Starbuck, A. L. Lentine, V. Stenger and S. Mookherjea, "Lightwave Circuits in Lithium Niobate through Hybrid Waveguides with Silicon Photonics," *Scientific Reports,* 2016.

[21] P. O. Weigel and S. Mookherjea, "Process Design Kit and Modulator Simulation for Hybrid Silicon-Lithium Niobate Integrated Optics," in *CLEO*, San Jose, 2017.

[22] D. M. Pozar, Microwave Engineering, Hoboken, NJ: Wiley, 2011.

[23] G. Ghione, Semiconductor Devices for High-Speed Optoelectronics, Cambridge: Cambridge University Press, 2009.

[24] G. E. Betts, "Microwave bandpass modulators in lithium niobate," *Integrated and Guided Wave Optics, 1989 Technical Digest Series,* vol. 4, pp. 14-17, 1989.

[25] Y. Shi, L. Yan and A. E. Willner, "High-Speed Electrooptic Modulator Characterization Using Optical Spectrum Analysis," *Journal of Lightwave Technologies,* vol. 21, no. 10, pp. 2358-2367, 2003.

[26] C. Hoessbacher, A. Josten, B. Baeuerle, Y. Fedoryshyn, H. Hettrich, Y. Salamin, W. Heni, C. Haffner, C. Kaiser, R. Schmid, D. L. Elder, D. Hillerkuss, M. Moller, L. R. Dalton and J. Leuthold, "Plasmonic modulator with >170 GHz bandwidth demonstrated at 100 GBd NRZ," *Optics Express,* vol. 25, pp. 1762-1768, 2017.

[27] Z. Ren, P. J. Heard, J. M. Marshall, P. A. Thomas and S. Yu, "Etching characteristics of LiNbO3 in reactive ion etching and inductively coupled plasma," *Journal of Applied Physics,* vol. 103, p. 034109, 2008.

[28] C. J. Kirkby, "Low-energy ion-beam processing damage in lithium niobate surface-acoustic-wave optical waveguide devices and its post-manufacture removal," *Journal of Materials Science,* vol. 27, no. 13, pp. 3637-3641, 1992.

[29] Y. K. D. L. E. S. W. H. Z. M. B. J. N. K. M. L. S. R. W. F. L. R. D. C. K. Clemens Kieninger, "Ultra-High Electro-Optic Activity Demonstrated in a Silicon-Organic Hybrid (SOH) Modulator," *ArXiv,* p. 1709.06338, 2018.

[30] C. Haffner, W. Heni, Y. Fedoryshyn, J. Niegemann, A. Melikyan, D. L. Elder, B. Baeuerle, Y. Salamin, A. Josten, U. Koch, C. Hoessbacher, F. Ducry, L. Juchili, A. Emboras, D. Hillerkuss, M. Kohl, L. R. Dalton, C. Hafner and J. Leuthold, "All-plasmonic Mach-Zehnder modulator enabling optical high-speed communication at the microscale," *Nature Photonics,* vol. 9, pp. 525-528, 2015.

[31] V. Sorianello, M. Midrio, G. Contestabile, I. Asselberg, J. Van Campenhout, C. G. Huyghebaerts, A. K. Ott, A. C. Ferrari and M. Romagnoli, "Graphene-silicon phase modulators with gigahertz bandwidth," *Nature Photonics,* vol. 12, pp. 40-44, 2018.

[32] E. L. Wooten, K. M. Kissa, A. Yi-Yan, E. J. Murphy, D. A. Lafaw, P. F. Hallemeier, D. Maack, D. V. Attanasio, D. J. Fritz, G. J. McBrien and D. E. Bossi, "A Review of Lithium Niobate Modulators for Fiber-Optic Communications Systems," *IEEE Journal of Selected Topics in Quantum Electronics,* 2000.

[33] P. A. Stolk, H.-J. Gossman, D. J. Eaglesham, D. C. Jacobson, C. S. Rafferty, G. H. Gilmer, M. Jaraiz, J. M. Poate, H. S. Luftman and T. E. Haynes, "Physical mechanisms of transient enhanced dopant diffusion in ion-implanted silicon," *Journal of Applied Physics,* vol. 81, pp. 6031-6050, 1997.

[34] M. M. Howerton and W. K. Burns, "Broadband traveling wave modulators in LiNbO3," in *RF Photonic Technology in Optical Fiber Links*, Cambridge University Press, 2002.

**Acknowledgments:**

P.O.W. acknowledges NDSEG fellowship support. Sandia is a multimission laboratory operated by National Technology and Engineering Solutions of Sandia, LLC, a wholly owned subsidiary of Honeywell International Inc., for the United States Department of Energy's National Nuclear Security Administration under contract DE-NA0003525. The views expressed in this article do not necessarily represent the views of the U.S. Department of Energy or the United States Government. The authors thank C. Levy and B. Rabet for useful discussions, as well as P. Asbeck for shared equipment.

**Funding:**

This work was partially supported by the San Diego Nanotechnology Infrastructure (NSF ECCS-1542148, UCSD Nano3 cleanroom) and through the United States Department of Energy's National Nuclear Security Administration under contract DE-NA0003525.

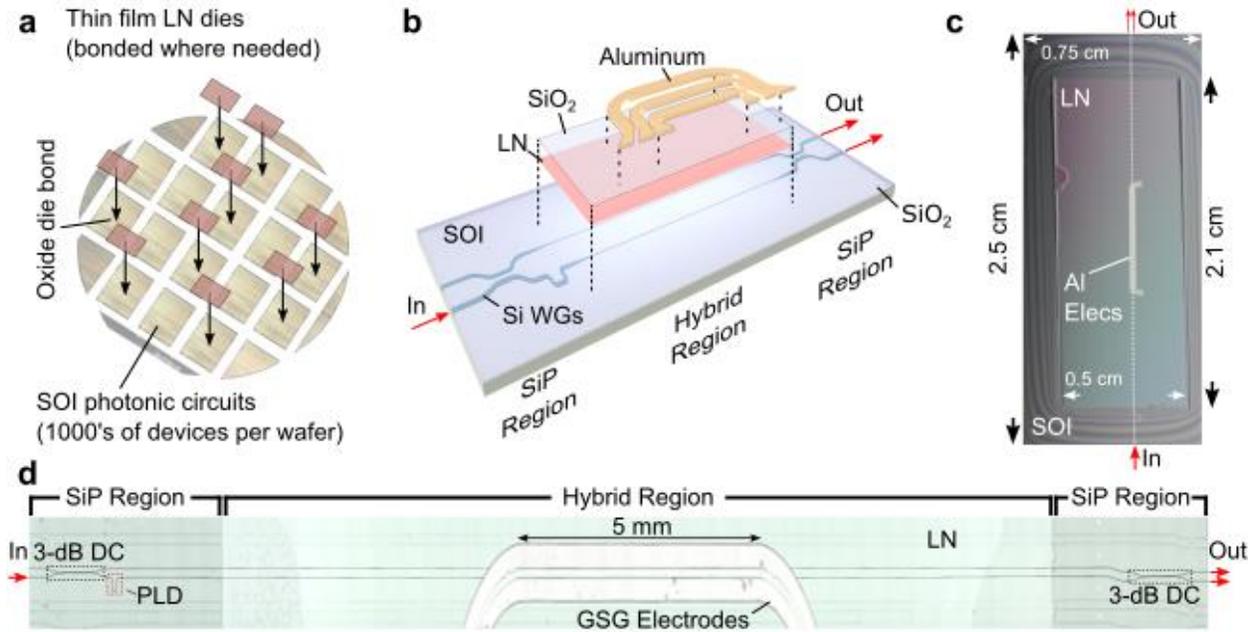

**Fig. 1a,** Thin film x-cut lithium-niobate (LN) on insulator dies were bonded at room temperature to segmented dies of a patterned and planarized silicon-on-insulator (SOI) wafer which contained fabricated silicon photonic waveguide circuits. No etching or patterning of the LN film was performed. **b**, Exploded representation of the EOM, where an unpatterned, un-etched LN thin film was bonded to a Mach-Zehnder interferometer fabricated in Si. Aluminum electrodes were deposited on a 50 nm $SiO_2$ layer over the LN film. 'SiP Region' denotes the $SiO_2$-clad region outside the bonded LN film, containing Si waveguide circuits, such as feeder waveguides, bends, directional couplers, and path-length difference segments. **c,** Top view of a representative fabricated hybrid Si-LN EOM test chip, which contains 60 EOM waveguide structures in parallel (in the north-south direction); for this report, test electrodes for use in push-pull configuration were only fabricated on one EOM device. **d,** Composite microscope image of the EOM. DC: directional coupler, PLD: path-length difference, GSG: ground-signal-ground, SiP: Si photonics.

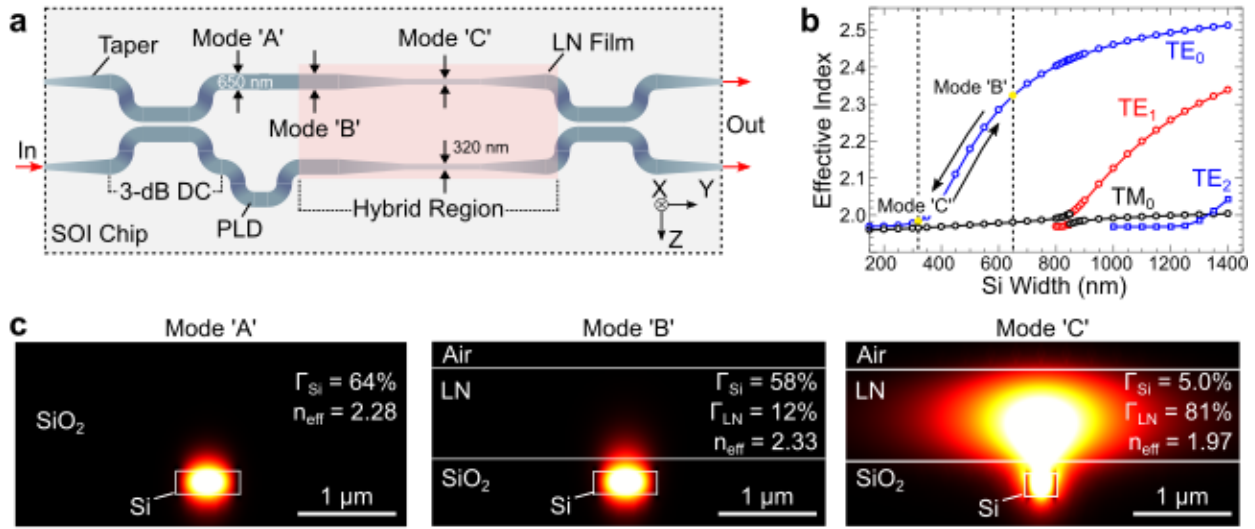

**Fig. 2a,** Schematic of the EOM (not to scale, not showing electrodes), including two 3-dB directional couplers (DC) and a waveguide segment for path-length difference (PLD). Three optical waveguide modes are used, labeled as A, B, and C. Modes A (Si under $SiO_2$) and B (Si under LN) have Si rib width w = 650 nm whereas mode C has w = 320 nm. **b,** Dispersion curves (effective index versus w) in the hybrid region; w values for modes B and C are chosen to stay within the single-mode region of operation. An adiabatic waveguide transition (variation in w) is designed to evolve from mode B to C and vice versa. **c,** Calculated Poynting vector components along the direction of propagation. Modes A and B are Si-guided and have a similar confinement fraction in Si. Mode C, with LN confinement factor ($\Gamma_{LN}$) greater than 80%, is used in the phase-shifter segments.

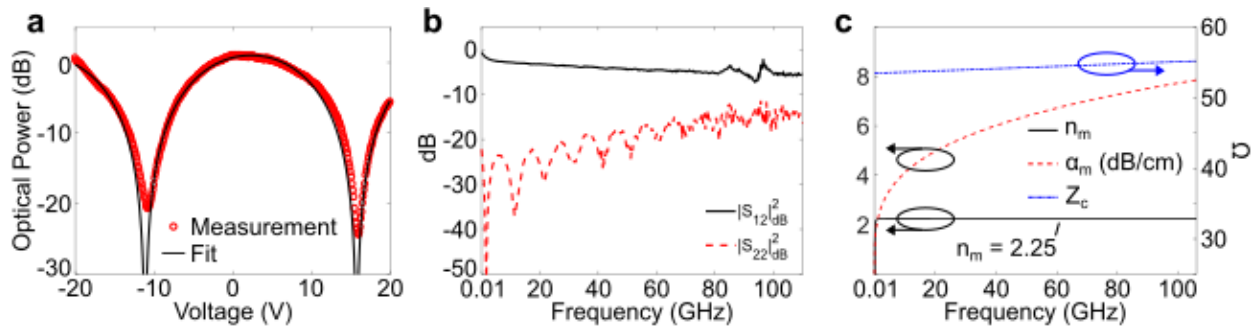

**Fig. 3a**, Normalized optical transmission of the Mach-Zehnder interferometric electro-optic Modulator (MZM), versus dc voltage at optical wavelength $\lambda$ = 1560 nm. Fitted $V_\pi L$ = 6.7 V.cm for device length L = 0.5

cm. **b,** Measured electrical S-parameters of the MZM's coplanar-waveguide transmission line. **c,** Left y-axis: extracted microwave phase index $n_m$ and microwave loss $\alpha_m$ (dB/cm) over the dc-110 GHz frequency range. Right y-axis: characteristic electrode impedance $Z_c$ ($\Omega$).

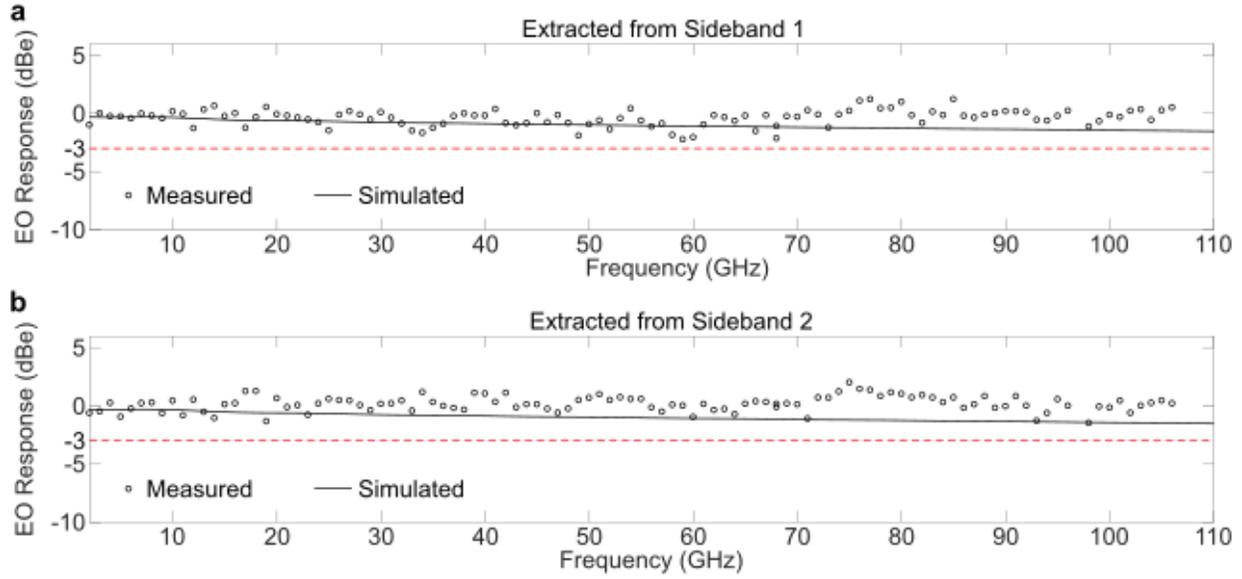

Fig. 4 Electro-optic response of the EOM for both sidebands (**a** and **b**) from the optical spectrum analyzer. Solid black line: calculated response from electrical S-parameters of Fig. 3c; black circles: electro-optic response from sideband OSA measurements.